\documentclass{nature}
\usepackage{graphics,epsf,epsfig,graphicx}

\title{Carbon Monoxide Clouds at Low-Metallicity in the WLM Galaxy}

\author{Bruce G. Elmegreen$^{1}$, Monica Rubio$^2$,
Deidre A. Hunter$^3$, Celia Verdugo$^{2}$, Elias Brinks$^4$, \&
Andreas Schruba$^5$}

\begin{document}

\maketitle

\begin{affiliations}
 \item IBM Research Division, T.J. Watson Research Center, 1101 Kitchawan
Road, Yorktown Heights, NY 10598, USA
 \item Departamento de Astronom\'ia, Universidad de Chile, Casilla 36-D, Santiago, Chile
 \item Lowell Observatory, 1400 West Mars Hill Road, Flagstaff, Arizona 86001 USA
 \item Centre for Astrophysics Research, University of Hertfordshire, Hatfield AL10 9AB, UK
 \item Cahill Center for Astronomy and Astrophysics, California Institute of
Technology, Pasadena, CA 91125, USA

\end{affiliations}

\begin{abstract}
Carbon monoxide (CO) is the primary tracer for interstellar clouds where
stars form, yet CO has never been detected in galaxies with an oxygen
abundance relative to hydrogen less than 20\% of solar, even though such low
metallicity galaxies often form stars. This raises the question of whether
stars can form in dense gas without the usual molecules, cooling to the
required near-zero temperatures by atomic transitions and dust radiation
rather than molecular line emission\cite{krumholz12}, and it highlights
uncertainties about star formation in the early universe, when the
metallicity was generally low. Here we report the detection of CO in two
regions of a local dwarf irregular galaxy, WLM, where the metallicity is 13\%
of the solar value\cite{lee05,asplund09}. New sub-millimeter observations and
archival far-infrared observations are used to estimate the cloud masses,
which are both slightly larger than $10^5\;M_\odot$. The clouds have produced
stars at a rate per molecule equal to 10\% of that in the local Orion Nebula
cloud. The CO fraction of the molecular gas is also low, about 3\% of the
Milky Way value. These results suggest that both star-forming cores and CO
molecules become increasingly rare in H$_2$ clouds as the metallicity
decreases in small galaxies.
\end{abstract}

Wolf-Lundmark-Melotte (WLM) is an isolated dwarf galaxy at the edge of the
Local Group\cite{leaman12}.  It has a low star formation rate because of its
small size, and like other dwarf Irregulars (dIrr), no previous
evidence\cite{tk01} for the molecular gas that always accompanies young stars
in larger galaxies\cite{bigiel11}. One problem is that the dominant tracer of
such gas is carbon monoxide (CO), and dIrr galaxies have low carbon and
oxygen abundances relative to hydrogen. No galaxy with $O/H$ abundance less
than 20\% has been detected in CO\cite{taylor98,leroy11,schruba12}. Far more
abundant is H$_2$, but this does not have an observable state of excitation
at the low temperatures ($\sim10-30$ K) required for star formation.

To search for star-forming gas, we surveyed WLM for CO($J = 3-2$) and
continuum dust emission at 345\,GHz using the Atacama Pathfinder EXperiment
(APEX) telescope at Llano de Chajnantor, Chile, with the the Swedish
Heterodyne Facility Instrument (SHFI\cite{vassilev}) and the Large APEX
Bolometer Camera (LABOCA\cite{siringo}). We also used a map of dust emission
at $160\,\mu$m from the {\em Spitzer} Local Volume Legacy
Survey\cite{dale09}, and a map of atomic hydrogen re-reduced from the
archives of the Jansky Very Large Array radio telescope. The dust
measurements convert to a dust temperature and dust mass, and after applying
a suitable gas-to-dust ratio, convert to a gas mass from which the HI mass
can be subtracted to give the H$_2$ mass for comparison to CO.

Figure 1 shows WLM and the two regions where we detected CO(3-2) emission,
along with HI, far-infrared and sub-mm images.  Observed and derived
parameters are in Tables 1 and 2, respectively. The peak CO brightness
temperature in each detected region is $\sim0.01$ -- 0.015 K and the FWHM
linewidth is $\sim12$ km s$^{-1}$. Previous efforts to detect CO(1-0) in
WLM\cite{tk01} partially overlapped Region A with a 45" aperture and
concluded a $5\sigma$ upper limit to the CO(1-0) intensity of 0.18 K km
s$^{-1}$. Our observation with an 18" aperture measures $0.200\pm0.046$ K km
s$^{-1}$ for CO(3-2) in the same region. The difference arises because the CO
cloud is unresolved even by our 18" beam -- we did not detect comparable
CO(3-2) intensities in our searches adjacent to Region A. The previous upper
limit corresponds to a maximum CO(1-0) luminosity of 8300 K km s$^{-1}$
pc$^{2}$ inside 45" (which is 215 pc at WLM), whereas the cloud we detect has
a CO(3-2) luminosity $\sim6$ times smaller, 1500 K km s$^{-1}$ pc$^{2}$.
Likewise, the previous null detection\cite{tk01} in CO(2-1) claimed a
$5\sigma$ upper limit that is about the same as our CO(3-2) detection, but
their closest pointing differed from our Region A by $\sim70$ pc (14" or half
the beam diameter at CO(2-1)), which could have been enough to take it off
the cloud.

The $160\,\mu$m, 870\,$\mu$m and HI peaks are slightly offset from the CO
positions, indicating variations in temperature and molecular fraction. A
large HI and FIR cloud that surrounds Region A, designated A1, was used to
measure dust temperature, $T_{\rm d}\sim15$K, which was assumed to apply
throughout the region (the $160\,\mu$m observation does not resolve Region A
so a more localized temperature measurement is not possible). $T_{\rm d}$ was
determined from the $870\,\mu$m and $160\,\mu$m fluxes corrected for the
CO(3-2) line and broadband free-free emission (Table 1) assuming a modified
black body function with dust emissivity proportional to the power $\beta$ of
frequency. Local measurements\cite{planck} suggest $\beta=1.78\pm 0.08$,
although a range is possible\cite{draine03,galametz12}, depending on grain
temperature and properties\cite{coupeaud11}. The $870\,\mu$m flux was also
corrected for an unexplained FIR and sub-mm excess that is commonly observed
in other low metallicity galaxies\cite{galametz11,verdugo12}. An alternate
combination of lower $\beta$ with no sub-mm correction\cite{plan11} gives
similar results (suppl. material). The dust mass was calculated from an
emissivity\cite{draine03} of $\kappa=13.9$ cm$^2$ gm$^{-1}$ at $140\,\mu$m,
and converted to $870\,\mu$m with the same power law index $\beta$. The dust
mass for Region A is then $M_{dust,A}=S_{\rm 870,A}D^2/\kappa B_{\nu}(T_{\rm
d})$ for flux $S_{\rm 870,A}$, distance $D$, and Black-Body function
$B_{\nu}.$

Dust mass is converted to gas mass with a factor equal to the gas-to-dust
mass ratio, $R_{\rm GD}$.  An approximation\cite{leroy11} is to assume the
solar value\cite{draine07} ($1/0.007$) increased by the inverse of the
metallicity of WLM ($0.13$), which would give $1100$. We use this
approximation here, but allow for a scaling factor to the gas mass,
$\delta_{\rm GD}=R_{\rm GD}/1100$; i.e., the gas-to-dust ratio normalized to
the Milky Way and scaled to the metallicity.

The results are in Table 2 assuming $\beta=1.8$ as the fiducial value, and
also $\beta=1.6$ and 2 to illustrate the dependence of the results on
$\beta$. Dust and gas mass correlate\cite{galametz12} with the assumed
$\beta$. The total gas mass column density in a 22" region around position A
is $\sim 58\pm15\;M_\odot$ pc$^{-2}$ for $\beta=1.8$. Atomic hydrogen
contributes $\sim27.3\pm1.4\;M_\odot$ pc$^{-2}$, and the remainder is
ascribed to molecular H$_2$ traced by the observed CO.

The integral under the CO(3-2) line from Cloud A is $I_{\rm
CO}=0.200\pm0.046$ K km s$^{-1}$. This has to be converted to CO(1-0) before
comparing it with H$_2$ mass in the conventional way.  We take as a
guide\cite{nikolic} the CO(3-2)/CO(1-0)$\sim0.80$ ratio in another low
metallicity galaxy, the Small Magellanic Cloud\cite{dufour84} (SMC, where
O/H$=20$\%). The result, 0.25 K km s$^{-1}$, is combined with the H$_2$ mass
column density to determine the conversion factor, $\alpha_{\rm CO}$, from
CO(1-0) to H$_2$. If $\alpha_{\rm CO}$ can be calibrated as a function of
metallicity, then CO observations can be used directly to infer the molecular
gas content irrespective of the dust spectral energy distribution. Extensive
compilations\cite{taylor98,leroy11,schruba12} show $\alpha_{\rm CO}$
increasing strongly at lower metallicity, from $\sim4\;M_\odot\;{\rm
pc}^{-2}\left( \rm{K \;km\; s}^{-1}\right)^{-1}$ in the Milky
Way\cite{leroy11} where the metallicity\cite{asplund09} is
$12+\log(O/H)=8.69$, down to the previous CO detection limit\cite{leroy11} in
the SMC where $\alpha_{\rm CO}\sim70\;M_\odot\;{\rm pc}^{-2}\left( \rm{K
\;km\; s}^{-1}\right)^{-1}$ at $12+\log(O/H)=8.0$. Our observations of
WLM\cite{lee05} at $12+\log(O/H)=7.8$ continue this trend.

Taking the H$_2$ column density from the residual between the dust-derived
total and the HI column density, $31\pm15\;M_\odot$ pc$^{-2}$, and dividing
by the inferred CO(1-0) line integral of 0.25 K km s$^{-1}$, we obtain
$\alpha_{\rm CO}= 124\pm60\;M_\odot\;{\rm pc}^{-2}\left( \rm{K \;km\;
s}^{-1}\right)^{-1}$ including Helium and heavy elements, with a range in
$\alpha_{\rm CO}$ from 34 to 271 as $\beta$ varies from 1.6 to 2. The
corresponding $X_{\rm CO}$ factor for H$_2$ column density conversion from
$I_{\rm CO}$ would be $(5.8\pm2.8)\times10^{21}$ cm$^{-2}$ $\left({\rm K\;
km\ s}^{-1}\right)^{-1}$, with a range from $1.5\times10^{21}$ to
$1.3\times10^{22}$ with $\beta$. There is a large uncertainty because of the
unknown dust properties ($\beta$, $\kappa$, $\delta_{\rm GD}$, sub-mm excess)
and molecular excitation (CO(3-2)/CO(1-0)) in dIrr galaxies.

The star formation rate based on the H$\alpha$ and FUV\cite{hunter10} fluxes
within an 18" diameter aperture centered on Cloud A is
$3.9-4.8\times10^{-5}\;M_\odot$ yr$^{-1}$. Dividing these rates into the
CO-associated molecular mass with $\alpha=124\;M_\odot$ pc$^{-2}$ (K km
s$^{-1})^{-1}$ gives a CO-molecular consumption time of 4.6-3.8 Gyr for
region A.  In Region B, the star formation rates by these two tracers are
$1.7-12.6\times10^{-5}\;M_\odot$ yr$^{-1}$ and the CO-molecular consumption
time is 6.7-1.5 Gyr. These times are only slightly larger than the average
value in spiral galaxies\cite{leroy08}, which is $\sim2$ Gyr, but they are
$10\times$ larger than the rate per molecule in local giant molecular
clouds\cite{lada12}, which is a more direct analogy with our observations.

The detection of CO in WLM suggests that star formation continues to occur in
dense molecular gas even at lower metallicities than previously observed. The
similarity between the metallicities of dIrr galaxies like WLM and those of
larger galaxies at high redshift\cite{mannucci09} implies that we should be
able to study star formation in young galaxies using the usual techniques.

\begin{addendum}
\item This work was funded in part by the National Science Foundation
    through grants AST-0707563 and AST-0707426 to DAH and BGE. MR and CV
    wish to acknowledge support from CONICYT (FONDECYT grant No 1080335). MR
    was also supported by the Chilean {\it Center for Astrophysics} FONDAP
    grant No 15010003. AS was supported by the Deutsche Forschungsgemeinschaft
    (DFG) Priority Program 1177.
    We are grateful to Marcus Albrecht for help with the
    LABOCA data reduction and to Lauren Hill for making Figure 1a.
    The National Radio Astronomy Observatory is a
    facility of the National Science Foundation operated under cooperative
    agreement by Associated Universities, Inc.

\item[Author Contributions] BGE coordinated the observational team,
    did the calculations for Table 2, and wrote the manuscript; MR did the telescope time
    justification and with CV observed
    the galaxy in CO and $870\,\mu$m using the APEX telescope with Chilean observing
    time, reduced the relevant
    data in Table 1, and also did relevant calculations for Table 2;
    DH determined the observational strategy, selected WLM for
    study, chose the observing coordinates, extracted the HI spectra
    from the LITTLE THINGS data cube,
    and prepared the figure. EB observed the galaxy on APEX with ESO
    time and coordinated the work on data uncertainties and background noise.
    AS made the WLM HI data cube from JVLA observations.  All authors discussed the
    results and commented on the manuscript.

Correspondence and requests for materials should be addressed to
bge@us.ibm.com

\end{addendum}

\begin{table}
\caption{{\bf Observations of WLM}}\label{table1}
\begin{tabular}{llllll}
\hline
Source & Reg. & RA & Dec & Beam (") & Flux \\
\hline
CO(3-2)   & A & 00:01:57.32 & -15:26:49.5 & 18     & $0.200\pm0.046$ K km s$^{-1}$ \\
HI        & A & 00:01:57.32 & -15:26:49.5 & 22  & $774\pm40$ mJy km s$^{-1}$\\
870 $\mu$m & A & 00:01:57.32 & -15:26:49.5 & 22   & $2.66\pm0.53$ mJy (0.11,0.02)$^a$ \\

HI        & A1 & 00:01:56.93 & -15:26:40.84 & 45  & $4170\pm82$ mJy km s$^{-1}$\\
870 $\mu$m & A1 & 00:01:56.93 & -15:26:40.84 & 45  & $15.2\pm 3.0$ mJy (0.11, 0.06)$^a$ \\
160 $\mu$m & A1 & 00:01:56.93 & -15:26:40.84 & 45  & $136.2\pm 13.6$ mJy (0.05)$^b$ \\

CO(3-2)   & B & 00:02:1.68 & -15:27:52.5 & 18    & $0.129\pm0.032$ K km s$^{-1}$ \\

\hline
\end{tabular}
$^a$ Quantities in parentheses are the CO(3-2) flux and the free-free
emission, both in mJy, that were subtracted from the source flux before
calculating the dust flux. $^b$ Quantity in parentheses is the free-free
emission, in mJy, that was subtracted from the source flux before calculating
the dust flux. The average FIR excess factor\cite{verdugo12} for the SMC is
1.7, so we divide the CO-corrected and free-free corrected $870\,\mu$m fluxes
in the table by 1.7 to get the thermal dust flux.
\end{table}

\begin{table}
\caption{{\bf Derived Quantities for WLM}}\label{table2}
\begin{tabular}{lllll}
\hline
Source & Reg. & $T$, K & $\Sigma$, $M_\odot$ pc$^{-2}$ & Mass, $M_\odot$\\
\hline
$\beta=1.8$, $\alpha_{\rm CO}=124\pm60^a$ &&&&\\
Dust      &A & $14.7\pm0.7^b$ &$0.053\pm0.014$ &$4.6\pm1.2\times10^2$ \\
Gas$^c$   &A &   &$(58\pm15)\delta_{\rm GD}$   &$(5.1\pm1.3\times10^5)\delta_{\rm GD}$ \\
HI$^d$    &A &   &$27.3\pm1.4$                 &$2.4\pm0.1\times10^5$\\
H$_2$     &A &   &$31\pm15$                 &$1.8\pm0.8\times10^5$\\
H$_2^e$     &B &   &$20\pm10$                 &$1.2\pm0.6\times10^5$\\

$\beta=1.6$, $\alpha_{\rm CO}=34\pm34^a$ &&&&\\
Dust      &A & $15.9\pm0.8^b$ &$0.032\pm0.008$ &$2.8\pm0.7\times10^2$ \\
Gas$^c$   &A &   &$(36\pm9)\delta_{\rm GD}$   &$(3.1\pm0.8\times10^5)\delta_{\rm GD}$ \\
H$_2$     &A &   &$8.3\pm9$                 &$0.5\pm0.5\times10^5$\\
H$_2^e$     &B &   &$5.3\pm6$                 &$0.3\pm0.3\times10^5$\\

$\beta=2$, $\alpha_{\rm CO}=271\pm97^a$ &&&&\\
Dust      &A & $13.6\pm0.6^b$ &$0.087\pm0.022$ &$7.5\pm1.9\times10^2$ \\
Gas$^c$   &A &   &$(95\pm24)\delta_{\rm GD}$   &$(8.3\pm 2.1\times10^5)\delta_{\rm GD}$ \\
H$_2$     &A &   &$67\pm24$                 &$3.9\pm1.4\times10^5$\\
H$_2^e$     &B &   &$44\pm16$                 &$2.5\pm0.9\times10^5$\\

\hline
\end{tabular}
\end{table}

\newpage

Table 2 notes: $^a$ The units of $\alpha_{\rm CO}$ are $\;M_\odot\;{\rm
pc}^{-2}\left( \rm{K \;km\; s}^{-1}\right)^{-1}$; the uncertainty is
dominated by the uncertainties in the $160\,\mu$m and $870\,\mu$m fluxes, as
indicated by their error limits in Table 1; the error limits are
approximately symmetric. $^b$ The dust temperature in region A is assumed to
be the same as the measured dust temperature in region A1. $^c$ $\delta_{\rm
GD}=R_{GD}/1100$ is the gas to dust ratio $R_{\rm GD}$ normalized to that of
the Milky Way scaled to the metallicity of WLM. Lowering $\delta_{\rm GD}$
lowers $\alpha_{\rm CO}$, but this does not seem reasonable: Engelbracht et
al.\cite{eng} show a correlation suggesting $M_{\rm gas}/M_{\rm
dust}\sim5000$ when $12+\log \mathrm{(O/H)}=7.8$, and this implies larger
$\delta_{\rm GD}\sim4.5$ and $\alpha_{\rm CO}$. The gas mass and resulting
$\alpha_{\rm CO}$ also depend on the assumed correction factor of 1.7 for
sub-mm excess. With no sub-mm excess correction, $\alpha_{\rm CO}$ increases
for all $\beta$: at $\beta=1.8$, $\alpha_{\rm CO}=370$. Solutions with no
sub-mm excess correction and lower $\beta$\cite{plan11} are in the
Supplementary material. In addition, $\alpha_{\rm CO}$ depends on the assumed
ratio CO(3-2)/CO(1-0), which was taken to be 0.8 for the table; a value of
CO(3-2)/CO(1-0)$=1$ increases $\alpha_{\rm CO}$ to 155 for $\beta=1.8$. $^d$
The HI mass column density is corrected for He and heavy elements.  $^e$ The
molecular mass for region B was calculated using the CO integrated intensity
and the value of $\alpha_{\rm CO}$ determined from Region A.

\clearpage

\begin{table}
\caption{{\bf (Alternate Method) Derived Quantities for WLM with no sub-mm Excess
Correction}}\label{table2}
\begin{tabular}{lllll}
\hline
Source & Reg. & $T$, K & $\Sigma$, $M_\odot$ pc$^{-2}$ & Mass, $M_\odot$\\
\hline
$\beta=1.2$, $\alpha_{\rm CO}=0.6\pm17$ &&&&\\
Dust      &A & $16.5\pm0.9$ &$0.025\pm0.007$ &$2.2\pm0.6\times10^2$ \\
Gas   &A &   &$(27\pm7)\delta_{\rm GD}$   &$(2.4\pm0.6\times10^5)\delta_{\rm GD}$ \\
HI    &A &   &$27.3\pm1.4$                 &$2.4\pm0.1\times10^5$\\
H$_2$     &A &   &$0.15\pm7$                 &$0.09\pm4.2\times10^4$\\
H$_2$     &B &   &$0.1\pm4.7$                 &$0.06\pm2.6\times10^4$\\

$\beta=1.4$, $\alpha_{\rm CO}=72\pm47$ &&&&\\
Dust      &A & $15.2\pm0.7$ &$0.041\pm0.011$ &$3.6\pm0.9\times10^2$ \\
Gas   &A &   &$(45\pm12)\delta_{\rm GD}$   &$(3.9\pm1.0\times10^5)\delta_{\rm GD}$ \\
H$_2$     &A &   &$18\pm12$                 &$1.0\pm0.7\times10^5$\\
H$_2$     &B &   &$11.6\pm7.5$                 &$6.7\pm4.4\times10^4$\\

$\beta=1.6$, $\alpha_{\rm CO}=186\pm75$ &&&&\\
Dust      &A & $14.1\pm0.6$ &$0.067\pm0.017$ &$5.8\pm1.5\times10^2$ \\
Gas   &A &   &$(74\pm19)\delta_{\rm GD}$   &$(6.4\pm 1.6\times10^5)\delta_{\rm GD}$ \\
H$_2$     &A &   &$46\pm19$                 &$2.7\pm1.1\times10^5$\\
H$_2$     &B &   &$30\pm12$                 &$1.7\pm0.7\times10^5$\\

\hline
\end{tabular}
\end{table}

\clearpage
\begin{figure}
\centering
\includegraphics[width=6.5in]{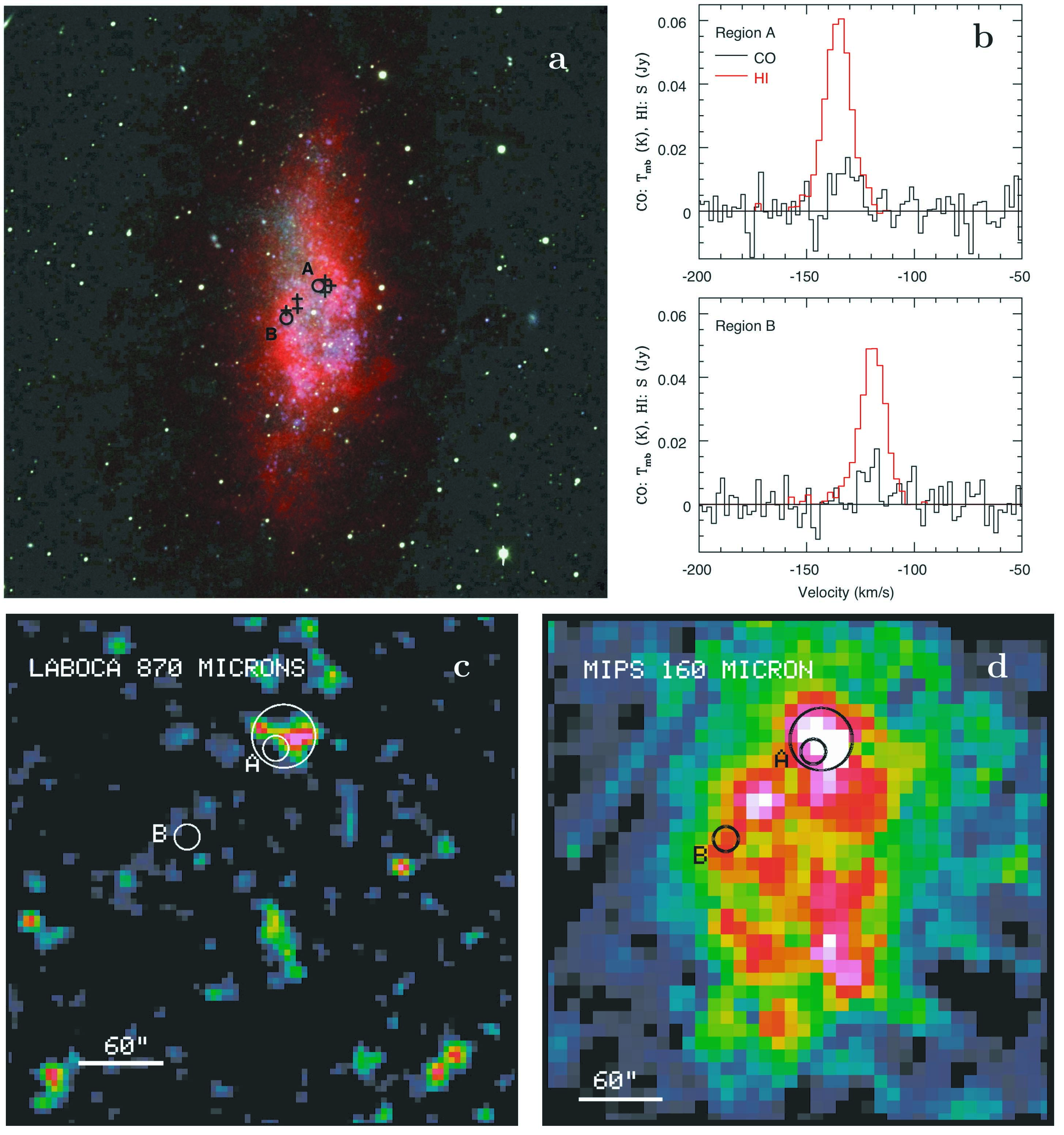}
\caption{WLM is a small, gas-rich galaxy $985\pm33$ kpc from the Milky Way
\cite{leaman12}. It contains $1.6\times10^7$ $M_\odot$ of stars\cite{zhang},
compared to $6.4\pm0.6\times 10^{10}\;M_\odot$ in the Milky
Way\cite{mcmillan11}, and it forms new stars at a rate\cite{hunter10} of 0.006 $M_\odot$
yr$^{-1}$, which is 12 times higher per unit stellar mass than
the Milky Way\cite{chomiuk11}. In the top-left color composite, red is HI,
green is V-band and blue is GALEX FUV. The HI has $7.6^{\prime\prime}$ and
2.6 km s$^{-1}$ resolution, and the CO(3-2) has $18^{\prime\prime}$ (circles)
and 0.11 km s$^{-1}$ resolution, although the CO(3-2) spectra shown in the figure were
smoothed to 2.2 km s$^{-1}$ resolution. The CO detections are labeled;
their exposure times were 218 (Region A) and
248 (Region B) minutes. Other searched regions with factors of $\sim2$ to
$\sim6$ shorter exposure times are shown by plus marks; the presence of
comparable CO in some of these other regions cannot be ruled out. Upper
right: Spectra of the two detections: velocities are Local Standard of Rest;
CO is main beam brightness temperature $T_{mb}$ in units of Kelvin; HI is
flux in units of Jy. Lower left: False color image of the $870\,\mu$m
observations taken with LABOCA on APEX. Lower right: False color {\em
Spitzer} $160\,\mu$m image obtained from {\em Spitzer} archives. The
$870\,\mu$m and $160\,\mu$m images display the same field of view. In the
lower panels, the small circles show where CO was detected and are
displayed with $22^{\prime\prime}$ diameter, which is the resolution of LABOCA.
The large circle is $45^{\prime\prime}$ in diameter and surrounds
a large HI and FIR cloud (A1) where the dust temperature was measured.}
\end{figure}

\end{document}